\documentclass[manuscript]{emulateapj}
\usepackage{natbib}
\usepackage{color}
\usepackage{amsmath}

\newcommand{\lya}{Ly$\alpha$}
\newcommand{\ha}{H$\alpha$}

\newcommand{\hii}{H~{\small II}}

\newcommand{\civ}{C~{\small IV}}
\newcommand{\ciii}{C~{\small III}}
\newcommand{\cii}{C~{\small II}}
\newcommand{\feii}{Fe~{\small II}}
\newcommand{\mgii}{Mg~{\small II}}
\newcommand{\siii}{Si~{\small II}}
\newcommand{\siiii}{Si~{\small III}}

\newcommand{\kms}{km s$^{-1}$}

\shortauthors{Scarlata \& Panagia}

\begin{document}


\title{A Semi--Analytical Line Transfer (SALT) model to interpret the
  spectra of galaxy outflows}


\author{C. Scarlata\altaffilmark{1} and N. Panagia\altaffilmark{2,3,4}}

 \altaffiltext{1}{Minnesota Institute for Astrophysics, School of
   Physics and Astronomy, University of Minnesota, 316 Church str
 SE, Minneapolis, MN 55455,USA}
 \altaffiltext{2}{Space Telescope Science Institute, 3700 San Martin
   Drive, Baltimore, MD 21218, USA,  panagia@stsci.edu}
\altaffiltext{3} {INAF--NA, Osservatorio Astronomico di Capodimonte, Salita Moiariello 16, 80131 Naples, Italy}
\altaffiltext{4}{ Supernova Ltd, OYV \#131, Northsound Rd., Virgin
  Gorda VG1150, Virgin Islands, UK}

\begin{abstract}
  We present a Semi--Analytical Line Transfer model, SALT, to study
  the absorption and re--emission line profiles from expanding
  galactic envelopes. The envelopes are described as a superposition
  of shells with density and velocity varying with the distance from
  the center.  We adopt the Sobolev approximation to describe the
  interaction between the photons escaping from each shell and the
  remaining of the envelope.  We include the effect of multiple
  scatterings within each shell, properly accounting for the atomic
  structure of the scattering ions. We also account for the effect of
  a finite circular aperture on actual observations.  For equal
  geometries and density distributions, our models reproduce the main
  features of the profiles generated with more complicated transfer
  codes. Also, our SALT line profiles nicely reproduce the typical
  asymmetric resonant absorption line profiles observed in
  star—forming/starburst galaxies whereas these absorption profiles
  cannot be reproduced with thin shells moving at a fixed outflow
  velocity.  We show that scattered resonant emission fills in the
  resonant absorption profiles, with a strength that is different for
  each transition. Observationally, the effect of resonant filling
  depends on both the outflow geometry and the size of the outflow
  relative to the spectroscopic aperture.  Neglecting these effects
  will lead to incorrect values of gas covering fraction and column
  density.  When a fluorescent channel is available, the resonant
  profiles alone cannot be used to infer the presence of scattered
  re--emission. Conversely, the presence of emission lines of
  fluorescent transitions reveals that emission filling cannot be
  neglected.

 \end{abstract}


\keywords{galaxies: ISM --- ISM: structure}

\section{Introduction}

The mechanical and radiative energy (from massive star winds and
supernova explosions) injected in the interstellar medium (ISM) of
galaxies is expected to drive gas outflows around regions of active
star--formation. These outflows are indeed observed both on the scale
of individual \hii\ regions as well as on full-galaxy scales. Outflows
are currently invoked as the principal mechanism regulating the
galactic baryonic cycle (i.e., the balance between the gas accretion
rate and the star-formation rate) in state-of-the-art galaxy formation
models \citep[e.g.][]{oppenheimer2010,dave2011,lilly2013}. Whether or
not the observed outflows are actually able to do the job is, however,
still unclear. In fact, although outflow-regulated models are able to
broadly reproduce some of the fundamental correlations observed in
massive galaxies, recent studies have pointed out the existence of a
fundamental problem with the evolution of low mass
($M_*<10^{9.5}M_{\odot}$) galaxies \citep{weinmann2012}.  This
difficulty appears to indicate that crucial feedback processes are
modeled incorrectly, since it is precisely in this low-mass regime
that feedback-induced outflows are expected to have the strongest
impact.

\begin{figure*}[ht!]
   \centering
  \includegraphics[scale=0.8]{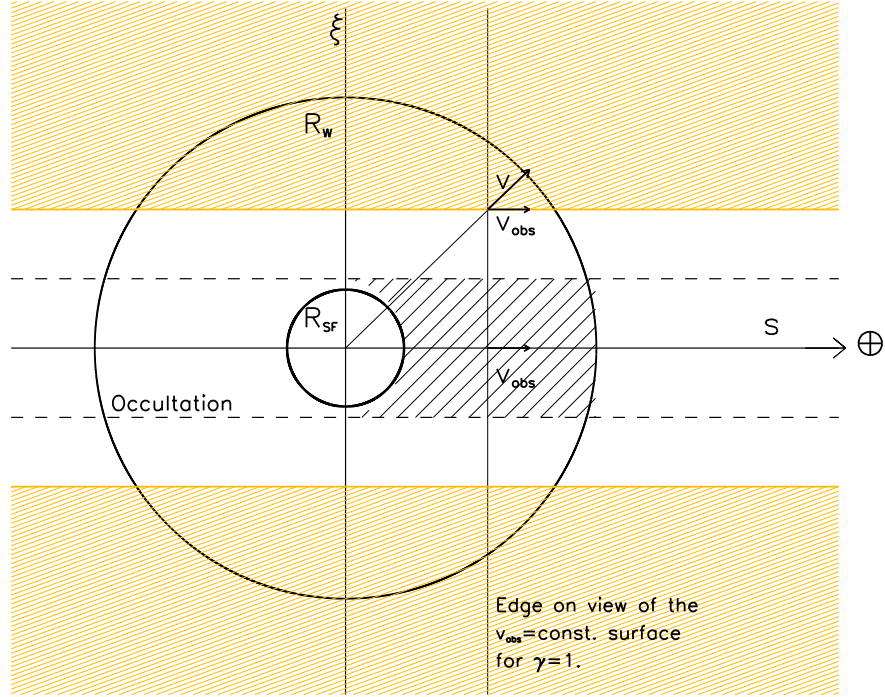}
  \caption{Side view of the outflow model. We consider a spherical
    galaxy (with radius $R_{\rm SF}$) surrounded by an expanding
    envelope with a velocity increasing radially with the distance
    from the center. For $r>R_W$ the gas velocity is $v_{\infty}$. The
    shaded yellow area shows the size of the aperture of radius
    $R_{\rm aper}$. The dotted vertical line indicates the edge-on
    view of the plane with constant observed velocity ($v_{\rm obs}$).
  }
   \label{fig:geometry}
\end{figure*}

What is lacking at this point are robust observational constraints on
the physical properties of galaxy outflows and how these depend on the
galaxy star--formation rates, stellar masses, and so on. In order to
characterize how effective feedback is in quenching the
star--formation (by, e.g., heating the gas and/or completely remove it
from a galaxy dark matter halo) we need to be able to probe the
kinematics of the outflows, their density structure, and their
extent. Absorption line studies against the strong UV continuum
produced in the star-forming regions can probe the neutral and ionized
components of the outflows, with numerous resonant transitions in low
and high ionization metals \citep[such as, \siii, \cii, \mgii, \ciii,
\feii, and\civ; e.g., ][]{rupke2005a,martin2005, sato2009,
  weiner2009,rubin2010,steidel1996, shapley2003}.

Recent studies show that outflowing gas moving with velocities up to
several hundreds of \kms\ is common in star--forming galaxies
\citep[e.g.,][]{pettini2002,martin2005}. Generally, these studies use
standard absorption--line analysis \citep[e.g][]{savage1991}, where
the velocity of the outflowing material is determined by the amount of
blueshift observed in the resonant absorption lines (typically UV
lines, but absorption of Na--D lines has been used). Recent
observations, however, have revealed the presence of numerous resonant
and fluorescent emission lines associated with the blueshifted
resonant absorptions \citep[e.g.,][]{weiner2009,
  france2010,rubin2011,jones2012, erb2012, martin2012,
  martin2013,kornei2013}. Although originally interpreted as the
result of photoionization by weak AGNs \citep[e.g.,][]{weiner2009},
these lines are now believed to be the result of scattered resonant
photons in an expanding envelope around galaxies
\citep[][]{rubin2011,erb2012,martin2013}.

The scattered re--emission into the line of sight affects the velocity
measurements based on pure absorption lines analysis, as well as mimic
a partial covering fraction of the outflowing gas
\citep[e.g.,][]{prochaska2011}. In these cases, it is crucial to be
able to consistently model both the resonant absorption as well as the
associated resonant and (if present) fluorescent emission originating
from the same ionic species. Steps in this direction have been taken by
\citet[][]{rubin2011}, who pioneered the study of outflows in emission
using resonantly-scattered MgII, and Fe lines.  Prochaska et
al. (2011) used Monte Carlo radiative transfer techniques to study the
nature of resonant absorption and emission for winds, accounting for
the effects of resonant scattering and fluorescence.

In this paper we go a step forward and develop a Semi--Analytical Line
Transfer (hereafter referred to as SALT) model to interpret the
absorption/scattered emission line profiles resulting from extended
galactic outflows. We assume that the Sobolev approximation holds, and
we account for multiple scattering within the outflow with a simple
statistical approach. As an example of an application of the model, we
apply SALT to multiple transitions in the $\rm Si^+$ ion observed in
the stack spectrum of $z\sim 0.3$ \lya\ emitters. We show how our
model is able to consistently reproduce the profiles of both the
absorption and resonant and fluorescent emission lines and thus
constrain the outflow velocity-- and density--fields.  Although rather
simplified, the SALT model can be used to gain a more physical
understanding of the outflowing gas, in both local and high redshift
galaxies.  We focus here on a few specific lines of Si$^+$ ion,
however, the models are readily applicable to any other ions.

The paper is organized as follows. In Section 2 we present the
derivation of the semi analytical line profile for an outflowing
spherical shell. The model is compared with the observed stacked
spectrum in Section~3, and the results are discussed in Section~4. We
offer our conclusions in Section~5.

\section{P-Cygni profile from a spherical expanding envelope}
\label{sec:model}
In spherical outflows, such as those produced in winds from early type
stars, resonant lines are characterized by the well understood
P--Cygni profile \citep[e.g.,][]{castor1970,castor1979}. The profile
shows both an emission and an absorption component. The absorption is
created in the material between the source and the observer, while the
emission is produced by scattering photons into the line of sight. For
spherical outflows, the velocity profile of the absorption component
will be blueshifted relative to the systemic velocity of the source
and will depend on the density and velocity fields of the absorbing
material. Because of the spherically symmetric geometry, the emission
component will be centered at the systemic velocity of the source and
thus will contribute to fill in the absorption at negative
velocities. This effect, well known in the context of stellar winds,
has generally been neglected in the context of absorption line studies
of galaxies. Recently, however, a few studies have emphasized the
importance of properly accounting for scattered re-emission in studies
of galaxy absorption line spectra. These studies have also highlighted
how the emission line features can be used as powerful diagnostics of
the geometry and physical conditions of galaxy outflows both in the
local and high--redshift Universe
\citep[][]{rubin2011,prochaska2011,erb2012, martin2013}.The goal of
this paper is to present a semi-analytical line transfer (SALT) model
that can be used to consistently interpret the absorption and emission
line profiles of both resonant and fluorescent transmissions observed
in galactic spectra. The line profiles are modeled for outflow
geometries similar to those introduced by the recent works of
\citet[][]{rubin2011,prochaska2011,erb2012, martin2013}.

In this Section we first describe the basic assumptions to derive the
line profiles in single scattering approximation \citet[following
][]{scuderi1992}. We then modify the wind model to include a more
realistic description of the scattering process by relaxing the single
scattering approximation. We also include the possibility of
re-emission in the fluorescent channel, different envelope geometries
and effect of a finite aperture.

\subsection{The basic model}
To build our SALT model we start with the simplest description for the
gas/star configuration. We approximate a galaxy as a spherical source
of UV radiation with radius $R_{SF}$ (i.e., where the bulk of
star-formation occurs), surrounded by an expanding envelope of gas,
extending to $R_{W}$.  In the following we will refer to the expanding
envelope as the ``galactic wind''.  Prochaska et al. (2011) use a
Monte Carlo radiative transfer technique to derive the absorption line
profiles resulting from similar galactic winds. In what follows, we
simplify their calculations by computing semi-analytical expressions
for the line profiles (including the effects of fluorescent emission
and spectroscopic aperture size). Martin et al. (2013) use a similar
outflow model to study the extended scattered emission from galactic
outflows at $z\sim 1$, considering the effects of a clumpy expanding
medium on the derived mass outflow rate.  Here, we build upon these
works, and we develop a semi-analytical algorithm to simply but
accurately calculate the expected line profiles originating in
extended outflows. Our results can be used to easily model galactic
extended outflows commonly observed in local and high redshift
galaxies.

In Figure~\ref{fig:geometry} we show the geometry of the wind and the
definition of the coordinate system. The coordinates $\xi$ and $s$ are
given in units of $R_{SF}$, so that in Figure~\ref{fig:geometry} the
dashed line tangential to the galaxy has $\xi=1$. We also introduce
the normalized radial coordinate $\varrho=r/R_{SF}$
($\varrho=\sqrt{\xi^2+s^2}/R_{SF}$).  A given point P in the envelope
is identified by the pair of coordinates $\varrho$, $\theta$, where
$\theta$ is the angle between the direction of $r$ and the line of
sight to the observer.

\begin{figure}[ht!]
   \centering
\includegraphics[scale=0.5]{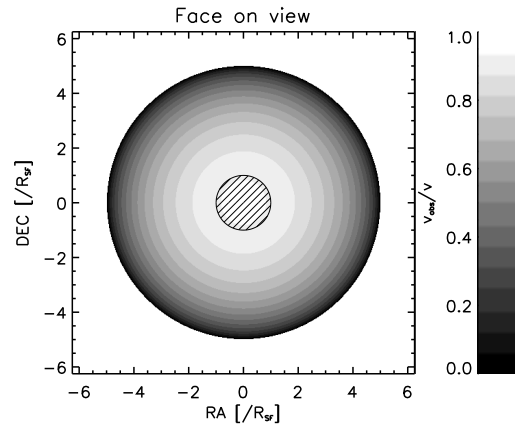}
 \caption{Iso-velocity contours of observed velocity for a shell
   moving at radial velocity $v$. The observed velocity ranges from
   $v_{\rm obs}=v$ at the center of the shell, to $v_{\rm obs}=0$ at
   the edge.}
   \label{fig:obs_velocity}
\end{figure}

We consider a velocity field where the velocity  ($v$) increases with
$r$ as a power law of exponent $\gamma$:

\begin{eqnarray}
v=v_0\, \Big( \frac{r}{R_{SF}} \Big)^{\gamma} \,\, {\rm for} \,\,  r \leq
R_{W}=R_{SF}\Big( \frac{v_{\infty}}{v_0} \Big) ^{ 1 / \gamma};
\end{eqnarray} 
 
\begin{eqnarray}
v=v_{\infty} \,\, {\rm for} \,\,   r \geq R_{W};
\end{eqnarray}
 
\noindent
where $v_0$ is the wind velocity at the surface of the star-forming
region (i.e., at $R_{SF}$), and $v_{\infty}$ is the terminal velocity
of the wind at $R_{W}$.

\begin{figure}
  \centering
\includegraphics[scale=0.2]{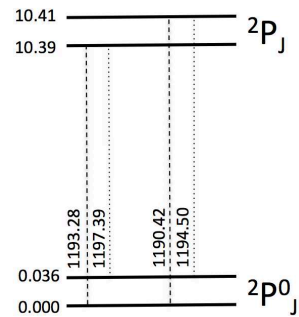}
 \caption{Energy levels of Si$^+$.}
   \label{fig:levels}
\end{figure}

When the velocity gradient in the expanding envelope is large, photons
will interact with the outflowing material only where the absorbing
ions are exactly ``at resonance'' due to their Doppler shift
\citep[this is a condition also known as ``Sobolev' approximation'',
e.g., ][]{sobolev}.  In this case, the radiative transport of the line
photons can be reduced to a local problem, and the optical depth for
absorption ($\tau$) can be evaluated at the {\it interaction surface},
which is defined in terms of the velocity as:

\begin{equation}
v= - c\frac{\Delta\,\nu}{\nu_0};
\end{equation}

\noindent
where $\nu_0$ is the resonance frequency of the line.  The numerical
results of Prochaska et al (2011) show that for the physical
conditions of typical galaxy outflows, the Sobolev approximation is a
justified assumption.

The wind optical depth at the interaction surface, can be written as a
function of wavelength, wind parameters, and atomic constants, as follows \citep[e.g.,][]{castor1970}:\\

\begin{eqnarray}
  \tau(r)=\frac{\pi e^2}{mc}f_{lu}\lambda_{lu}n_l(r)\Big[1-\frac{n_u g_l}{n_lg_u}\Big]\frac{r/v}{1+\sigma\mu^2};
 \end{eqnarray}

\noindent
where $f_{ul}$ and $\lambda_{lu}$ are the oscillator strength and
wavelength, respectively, for the $ul$ transition, $\mu=\cos(\theta)$,
and $\sigma=\frac{{\rm d}\ln(v)}{{\rm d}\ln(r)}-1$ (see
Table~\ref{tab:siII} for Si$^+$ atomic data).  The expression for the
optical depth can be simplified by assuming that 1) it does not depend
on the angle $\theta$ between a radius and the line of sight, 2) the
above velocity law holds, 3) stimulated emission is negligible (i.e.,
$\Big[1-\frac{n_u g_l}{n_lg_u}\Big]=1$), and 4) the mass outflow rate
is constant, so that $n_l(r)\propto (vr^2)^{-1}$.  For $\gamma=1$, and
with these assumptions, we can write:
 
\begin{eqnarray}
\tau(r)=\frac{\pi e^2}{mc}f_{lu}\lambda_{lu}n_{0}\left(\frac{R_{SF}}{r}\right)^3 \frac{r}{v}\\
=\tau_0 \left(\frac{R_{SF}}{r}\right)^3
\end{eqnarray}

and 

\begin{equation}
\tau_0= \frac{\pi e^2}{mc}f_{lu}\lambda_{lu}  n_{0}\frac{R_{SF}}{v_0};
\label{eqn:tau0}
\end{equation}

\noindent
where $n_0$ is the gas density at $R_{SF}$ (for $\gamma=1$,
$n_l(r)=n_{0}\Big(\frac{r}{R_{SF}}\Big)^{-3}$). 

Now consider a thin shell located at a distance
$r = R_{SF} (\frac{v}{v_0})^{1/\gamma}$, moving with an intrinsic
radial velocity $v$.  The velocity measured by the observer, i.e. the
component of the radial velocity along the line of sight to the
observer ($v_{\rm obs}=v\,\cos{\theta}$) will depend on the position
on the shell and in particular on the projected distance to the
center. This is shown in Figure~\ref{fig:obs_velocity}, where we plot
contours of constant {\it observed} (i.e., projected) velocity from a
shell moving outward with radial velocity $v$. For the sake of
clarity, we show only the half of the shell moving toward the
observer. The observed velocities range from $-v$ at the projected
center of the envelope (where the gas is moving directly toward us) to
0 at the projected distance $r=r_v$ (where the shell is moving on the
plane of the sky). Obviously, only the portion of the shell {\it in
  front} of the continuum disk (hatched area in
Figures~\ref{fig:obs_velocity}) will produce a net absorption in the
spectrum (blueward of the line center) by scattering photons out of
the line of sight.

Globally the shell will absorb a fraction $E(v)=[1-\exp{(-\tau(v))}]$
of the energy that, in terms of observed velocities
($v_{\rm obs}=v\,\cos{\theta}$), will be redistributed evenly over the
velocity interval ($v_{\rm min}$, $v$). Here $v_{\rm min}$ is the
projection of the shell velocity along the line of sight tangential to
the galaxy (i.e., at $\xi =1$). Following Scuderi {\it et al.} (1992),
we can compute $v_{min}$ as:

 \begin{equation}
v_{\rm min}=v\,\cos{\theta}=v\,\frac{s(\varrho(v))}{\varrho(v)}.
\end{equation}

\noindent
Or, setting $y=v/v_0$, as:

 \begin{equation}
y_{\rm min}=y^{(\gamma -1)/\gamma}\, (y^{2/\gamma}-1)^{1/2}.
\label{eqn:vmin}
\end{equation}

Only shells with intrinsic radial velocities in the range from $v_{\rm
  obs}$ and $v_1=v_{\rm obs}/\cos{\theta}$ (for $\xi = 1$) can
contribute to the absorption at $v_{\rm obs}$. Setting $x=v_{\rm
  obs}/v_0$, $y_1=v_1/v_0$ can be computed by solving the equation:

\begin{equation}
y_1^2\,(1-y_1^{-2/\gamma})=x^2.
\end{equation}

Thus, we can write the absorption component of the profile, in units
of the stellar continuum as:

 \begin{equation}
I_{\rm abs, blue}(x)=\int_{max(x,1)} ^{y_1} \, \frac{1 - e^{-\tau (y)}}{y-y_{\rm min}} \, \mathrm{d}y.
\end{equation}

Surfaces of constant {\it observed} velocity can be described by the
equation:

\begin{equation}
  v_{\rm obs}=v_0\left(\frac{r}{R_{SF}}\right)^{\gamma}\cos{\theta}.
\end{equation}

\noindent
For the particular case of $\gamma=1$, this equation describes
parallel planes at distance
$r=R_{SF}\frac{v_{\rm obs}}{v_0}\cos{\theta}$ from the center of the
emitting region (see Figure~\ref{fig:geometry}).

\subsubsection{Single scattering approximation}
Resonant photons absorbed in the envelope can be detected when
re--emitted toward the observer.  Assuming that a re-emitted photon
escapes a given shell without further interactions, we can compute the
emission component of the line profile as follows.  If the photons are
re--emitted isotropically, then they will uniformly cover the range of
projected velocities between $\pm v$. To describe the profile of
this emission component, we divide the range of observed velocities
into blueward and redward of the systemic velocity.  The blue side of
the emission profile originates in the half of the envelope
approaching the observer (i.e., $s\ge 0$). For a given observed
velocity, the emission will come from all shells with $v>v_{\rm obs}$,
and we can write:

\begin{equation}
I_{\rm em, blue}(x)=\int_{max(x,1)} ^{y_{\infty}} \, \frac{1 -
  e^{-\tau (y)}}{2y} \, \mathrm{d}y.
\label{eqn:iemblue}
\end{equation}

The red side of the profile is produced in the unocculted, receding
portion of the envelope. Because of the occultation, however, only
shells with velocities larger than $v_{\rm min}$ (see
Eq.~\ref{eqn:vmin}) will contribute to a given observed receding velocity:

\begin{equation}
I_{\rm em, red}(x)=\int_{y_1} ^{y_{\infty}} \, \frac{1 - e^{-\tau (y)}}{2y} \, \mathrm{d}y.
\label{eqn:red}
\end{equation}

Finally, the resulting P-Cyg profile for the ideal spherical outflows
can be computed as:

\begin{equation}
I(x)=1-I_{\rm abs,  blue}+I_{\rm em, blue}+I_{\rm em, red}.
\label{eqn:full}
\end{equation}

\begin{figure*}[t!]
   \centering
 \includegraphics[scale=0.5]{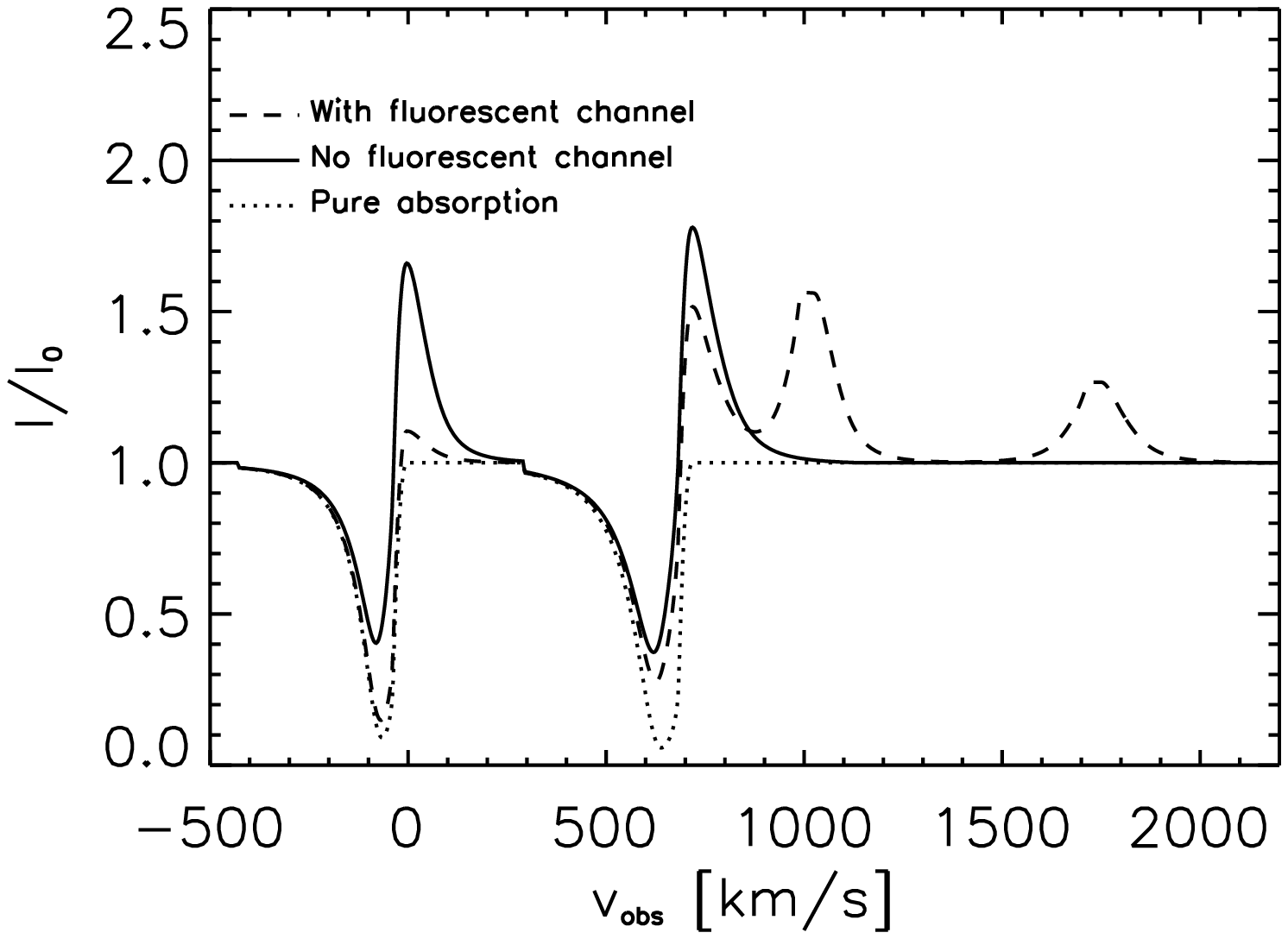}
 \includegraphics[scale=0.5]{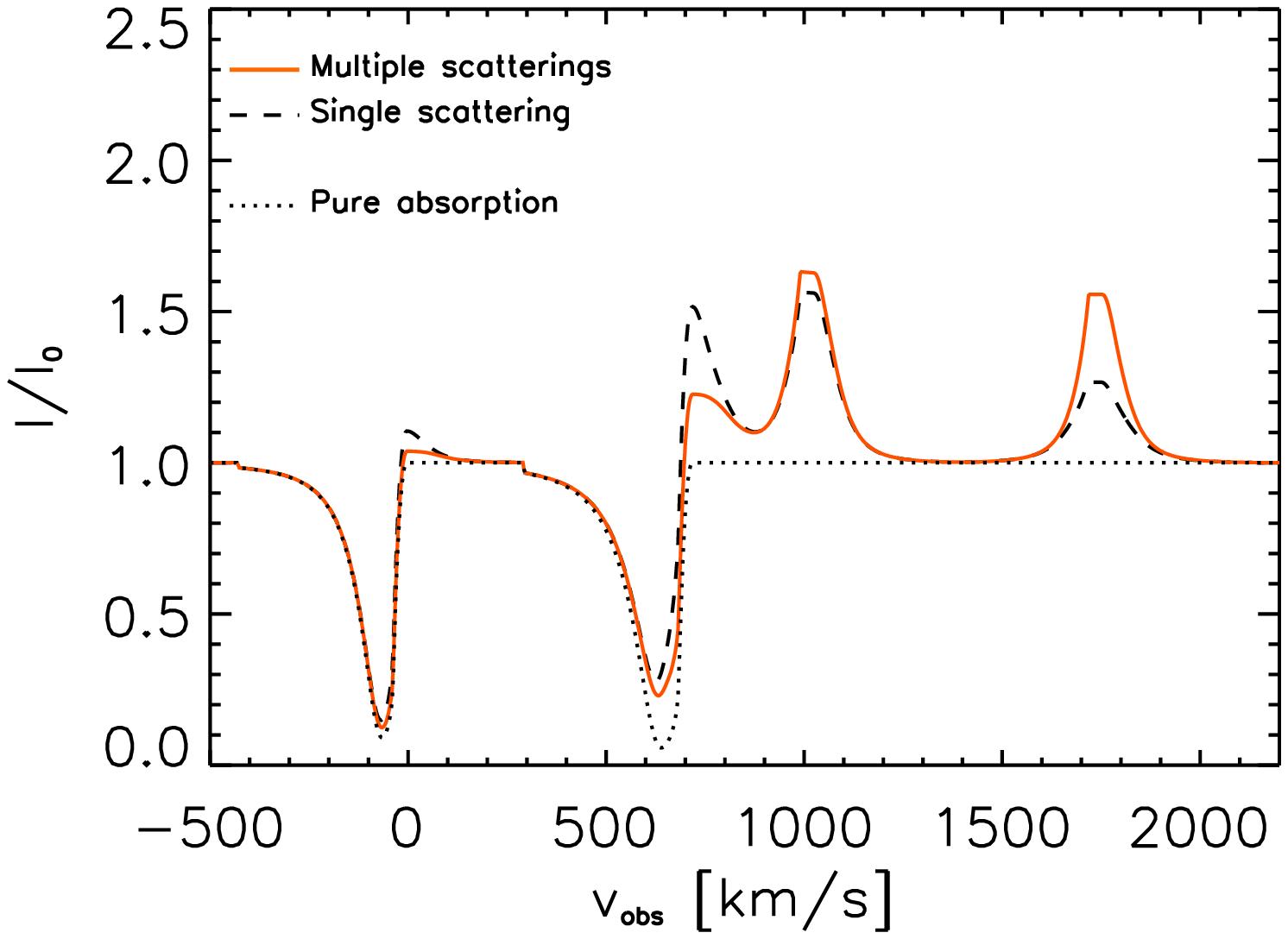}
 \caption{Left: Effect of fluorescent channel on the resonant P-Cygni
   profile. The Si$^+$ doublet line profiles are shown with and
   without the inclusion of the fluorescent emission (dashed and solid
   line respectively) for the single-scattering approximation. The
   pure absorption component of the profile is also shown for
   reference (dotted line). Right: Si$^+$ doublet profiles computed
   with single scattering approximation (black) and multiple
   scatterings (red).}
   \label{fig:fluo}
\end{figure*}

\subsubsection{Fluorescent emission in single scattering approximation}
Depending on the energy levels of the particular ionic species, the
absorption of a resonant photon can result in the production of a
fluorescent photon. This occurs when the electron decays into an
excited ground level\footnote{In what follows, fluorescent transitions
  will be indicated with an $*$.}. As an example,
figure~\ref{fig:levels} shows the energy level diagram of the
$\lambda$~1190.42 and 1193.28\AA\ $\rm Si^+$ doublet.  Resonant and
fluorescent transitions are marked with dashed and dotted lines,
respectively. We account for the fluorescent channel in the modeling
of the line profile as follows.

For a bound electron, the probability of decaying into the lower level
$l$ is proportional to $p_{ul}=A_{ul}/\sum_i{A_{ui}}$, where $A_{ui}$
is the spontaneous decay probability from the upper level $u$ to the
lower level $i$. Relevant $A_{ui}$ values are given in
Table~\ref{tab:siII}.  In the single scattering approximation, the
resulting line profile accounting for the emission in the fluorescent
channel becomes:

\begin{equation}
I(x)=1-I_{\rm abs,  blue}+p_{R}\, (I_{\rm em, blue}+I_{\rm em, red}) +p _{F} \, (I_{\rm em, blue}+I_{\rm em, red}) ,
\label{eqn:fullfluo}
\end{equation}

\noindent where $p_{R}$ and $p_{F}$ are the probabilities that a
photon is re--emitted in the resonant and the fluorescent channels,
respectively. When only one fluorescent channel is available, as in
the cases considered here, $p_R+p_F=1$. The left panel in
Figure~\ref{fig:fluo} shows how the line profiles of the \siii\
doublet generated in an outflowing envelope change when the
fluorescent channels are taken into account. When a fraction of the
photons are re--emitted in the fluorescent transition, the filling
effect of the resonant absorption due to photons scattered in the wind
is substantially reduced for the 1190\AA\ transition. It is only
minimally reduced for the transition at 1193\AA\ due to contamination
from the fluorescent re-emitted photons at 1194.5\AA\ (see
Table~\ref{tab:siII}). Because of the re--emission in the fluorescent
channel, the total equivalent width of the resonant P-Cygni profile
(i.e., including both the absorption and emission components) is
negative (net absorption).

\subsubsection{Accounting for multiple scatterings}

More realistically, a photon re-emitted with the resonant energy will
likely interact with the ions in the shell where it was created,
resulting in multiple scattering events of a single photon within a
given shell. Multiple scatterings will not change the shape of the
absorption profile ($I_{abs,blue}$), but will reduce the contribution
of the re--emission in the resonant line, while enhancing the
re--emission in the fluorescent channel (when this is available). The
number of scatterings will clearly be a function of the ion density at
any given point in the outflow. Thus, for our assumed density profile,
this process will be more important in the internal regions of the
outflow, where the density is highest. We account for multiple
scatterings within a single shell, as follows.

 We define the photon's escape probability from a shell of optical
  depth $\tau(v)$ as \citep[e.g.,][]{Mathis1972}:

\begin{equation}
  \beta=(1- e^{-\tau})/\tau, 
\end{equation}

\noindent
Thus, for a shell with velocity $v$, a photon has a probability
$\beta$ of escaping the shell, and therefore --because of the
underlying Sobolev approximation-- escaping the outflow.  Of all
photons absorbed at resonance by the moving shell, a fraction $p_F$
will be re-emitted in the fluorescent channel and escape. Of the
fraction $p_R$ of the photons re-emitted at resonance, a faction
$1-\beta$ will be absorbed again before they are able to escape the
shell.  Of these [$p_R\,(1-\beta)$], a fraction $p_F$ will be
converted into fluorescent photons and escape [i.e.,
$p_F\,p_R\,(1-\beta)$]. Again, out of the resonantly re--emitted
photons, a fraction $1-\beta$ will be re--absorbed within the shell,
contribute to the fluorescent re--emission and escape the outflow. It
can be easily shown that, {\it for each shell}, the fraction of
absorbed photons converted into fluorescent photons is given by:

\begin{equation}
F_{F}(\tau)=p_{F}\sum_{n=0}^{\infty} [p_R\,(1-\beta)]^n,
\label{eqn:msf}
\end{equation}

\noindent
while the fraction of absorbed photons that are able to escape will
be:

\begin{equation}
F_{R}(\tau)=p_{R}\beta\sum_{n=0}^{\infty} [p_R\,(1-\beta)]^n.
\label{eqn:msr}
\end{equation}

\noindent
Because $p_R\,(1-\beta)<1$, the summation of the geometric series in
Equations~\ref{eqn:msf} and \ref{eqn:msr} converges, and

\begin{subequations}
\begin{align}
F_F &=p_F/[1-p_R\,(1-\beta)] \\
F_R &=\beta\,p_R /[1-p_R\,(1-\beta)].
\end{align}
\end{subequations}

\begin{figure}[t!]
   \centering
  \includegraphics[scale=0.5]{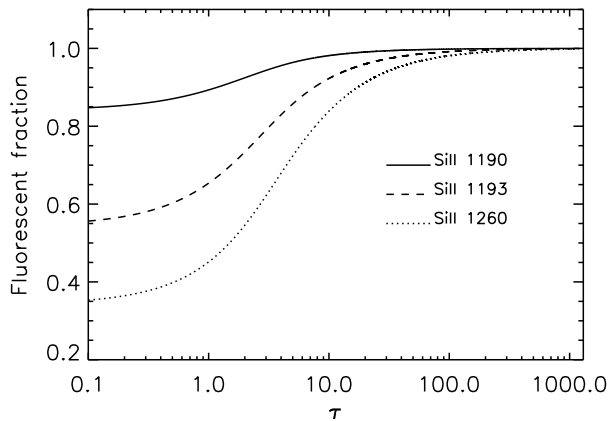}
  \caption{The fraction of absorbed resonant photons re-emitted in the
    fluorescent channel after multiple scatterings depends on the gas
    column density as well as on the transition probability. We show
    the calculations for three transitions of Si$^+$, as indicated in
    the label.}
   \label{fig:fluo_frac}
\end{figure}

Figure~\ref{fig:fluo_frac} shows the fraction of absorbed photons that
escape in the fluorescent channel as a function of the shell optical
depth for three representative values of $p_F$.  Analogously to
Eqns~\ref{eqn:iemblue} and \ref{eqn:red}, the blue and red
resonant-emission components become:

\begin{subequations}
\begin{align}
  I_{\rm em, blue,MS}(x)&=\int_{max(x,1)} ^{y_{\infty}} \, F_{R}(y)\frac{1 -
    e^{-\tau (y)}}{2y} \, \mathrm{d}y\\
I_{\rm em, red,MS}(x)&=\int_{y_1} ^{y_{\infty}} \, F_{R}(y)\frac{1 - e^{-\tau (y)}}{2y} \, \mathrm{d}y,
\label{eqn:redMS}
\end{align}
\end{subequations}

\noindent
while the blue and red fluorescent components can be written as:

\begin{subequations}
\begin{align}
  I_{\rm em, blue,MS,F}(x)&=\int_{max(x,1)} ^{y_{\infty}} \, F_{F}(y)\frac{1 -
    e^{-\tau (y)}}{2y} \, \mathrm{d}y\\
I_{\rm em, red,MS,F}(x)&=\int_{y_1} ^{y_{\infty}} \, F_{F}(y)\frac{1 - e^{-\tau (y)}}{2y} \, \mathrm{d}y.
\label{eqn:fullMS}
\end{align}
\end{subequations}

In the right panel of Figure~\ref{fig:fluo} we show the effect of
accounting for multiple scatterings on the line profiles of the SiII
doublet. As expected, the re-scattered emission component at resonance
is reduced significantly compared to the single-scattering
approximation resulting in an increased intensity of the fluorescent
lines. We note that this effect enhances the contamination of the
SiII~1193\AA\ absorption component from re--emitted fluorescent
photons from the SiII~1190\AA\ transmission. This enhancement is
particularly prevalent in low resolution spectra.

\begin{figure}[t!]
   \centering
  \includegraphics[scale=0.5]{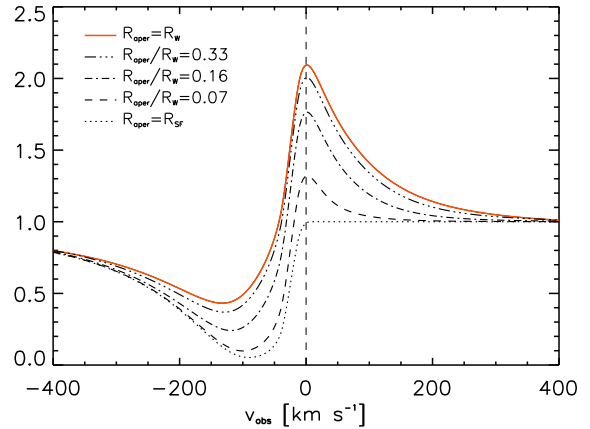}
  \caption{The P-Cygni line profile changes as function of size of
    spectroscopic aperture, from a pure absorption profile ($R_{\rm
      aper}=R_{\rm SF}$), to a classic P-Cyg profile ($R_{\rm
      aper}=R_{\rm W}$). The profiles where computed assuming a
    spherical expanding envelope, $\gamma=1$, $\tau=60$, $v_0=25$ \kms\
    and $v_{\infty}=450$ \kms\ (we consider here a line with no
    fluorescent transition, such as, e.g., SiIII $\lambda=1260$).}
   \label{fig:slit}
\end{figure}

\subsection{Spherical envelope observed with a circular finite aperture}
\label{sec:aperture}
If the spectroscopic observations are made using an aperture that does
not include the full extent of the scattering envelope, then the
observed line profile can change dramatically. To illustrate the
consequences, we consider here the case of a spherical envelope
observed with a circular aperture larger than the central source but
smaller than the entire envelope, i.e., with $R_{\rm aper}\ge R_{SF}$
and $R_{\rm aper} \le R_{W}$. We also restrict our analysis to $\gamma
=1$.

Clearly, the blueshifted absorption component of the profile will
remain unchanged due to the presence of the aperture. The blue and red
scattered emission, however, will change. In fact, for a shell of
intrinsic velocity $v$, the aperture will block those photons
scattered at velocities smaller than
$v_{\rm aper}=v\,cos(\theta_{\rm aper})$, where $\theta_{\rm aper}$ is
such that $sin(\theta_{\rm aper})= R_{\rm aper}/r_v$. Clearly, see
Figure~\ref{fig:geometry}, each shell will correspond to a different
$\theta_{\rm aper}$.

In Figure~\ref{fig:geometry} we show the edge-on view of a plane of
constant $v_{\rm obs}$ ({\it dotted line}). As we saw earlier, only
layers with $v \ge v_{\rm obs}$ will contribute to the emission at
$v_{\rm obs}$.  Figure~\ref{fig:geometry} shows that the effect of
adding a circular aperture is to remove the contribution at $v_{\rm
  obs}$ from all shells with $v>v_{\rm up}$, where:

\begin{equation}
v_{\rm up}=\frac{v_{\rm obs}}{cos\theta_{\rm aper}}.
\end{equation}

In terms of $v_0$, and after a little algebra, we get: 

\begin{equation}
y_{\rm up}^2=x^2+\left(\frac{R_{\rm aper}}{R_{SF}}\right)^2.
\end{equation}

Thus, $I_{em, blue}^{aper}(x)$ (see Eq.~\ref{eqn:iemblue}) can be
written here as:

\begin{equation}
I_{\rm em, blue}(x)^{aper}=\int_{max(x,1)} ^{y_{\rm up}} \, \frac{1 - e^{-\tau (y)}}{2y} \, \mathrm{d}y;
\end{equation}

Analogously, $I_{em, red}^{aper}(x)$ will be:

\begin{equation}
I_{\rm em, red}(x)^{aper}=\int_{y_1} ^{y_{\rm up}} \, \frac{1 - e^{-\tau (y)}}{2y} \, \mathrm{d}y;
\end{equation}

In Figure~\ref{fig:slit} we show how the P-Cygni profile changes with
the ratio $R_{\rm aper}/R_{SF}$. In the extreme case of $R_{\rm
  aper}/R_{SF}=1$ (dotted line), i.e., when the aperture is only as
large as the source of continuum, the line is observed only in
absorption, and blueshifted relative to the systemic velocity of the
galaxy. A component of scattered re--emitted photons coming from the
absorbing material is contributing to the blue side of the line, but
no re--emitted photons are detected on the red side, because of both
 the effect of the aperture, and the -often neglected- occultation by the
galaxy.

As the ratio $R_{\rm aper}/R_{SF}$ increases, fewer photons scattered
toward the observer are blocked by the aperture. As a result, the
emission component of the profile (centered at the systemic velocity)
becomes more and more pronounced. The shape of the absorption profile
also changes because of the increasing contribution of photons
scattered by material moving toward the observer. It is also evident
from Figure~\ref{fig:slit} that the velocity at maximum absorption
shifts toward higher blueshifted velocities as the the contribution
from scattered emission increases (i.e., as $R_{\rm aper}/R_{SF}$
increases).

\section{Application to real spectra}
\label{sec:comparison_observations}
As an example of its flexibility, we apply the SALT model to resonant
line profiles observed in a stacked spectrum of \lya\ emitting
galaxies. We first summarize the data and the measurements
(Section~\ref{sec:data}) and then discuss the properties of that
outflow that can be inferred from the absorption line analysis
performed with the SALT model.

\subsection{Data and measurements}
\label{sec:data}
The average spectrum modeled in this section was created by stacking
Cosmic Origin Spectrograph \citep[COS][]{COS} medium-resolution
spectra of a sample of 25 known $z\sim0.3$ \lya\ emitters
\citep{deharveng2008, cowie2010}. The details of the data reduction
and spectral extraction are presented in Scarlata et al. (2014). For
each galaxy, we have accurate redshift measurements obtained from the
\ha\ emission line profiles \citep{cowie2011}.

\begin{figure*}[h]
   \centering
  \includegraphics[scale=0.7]{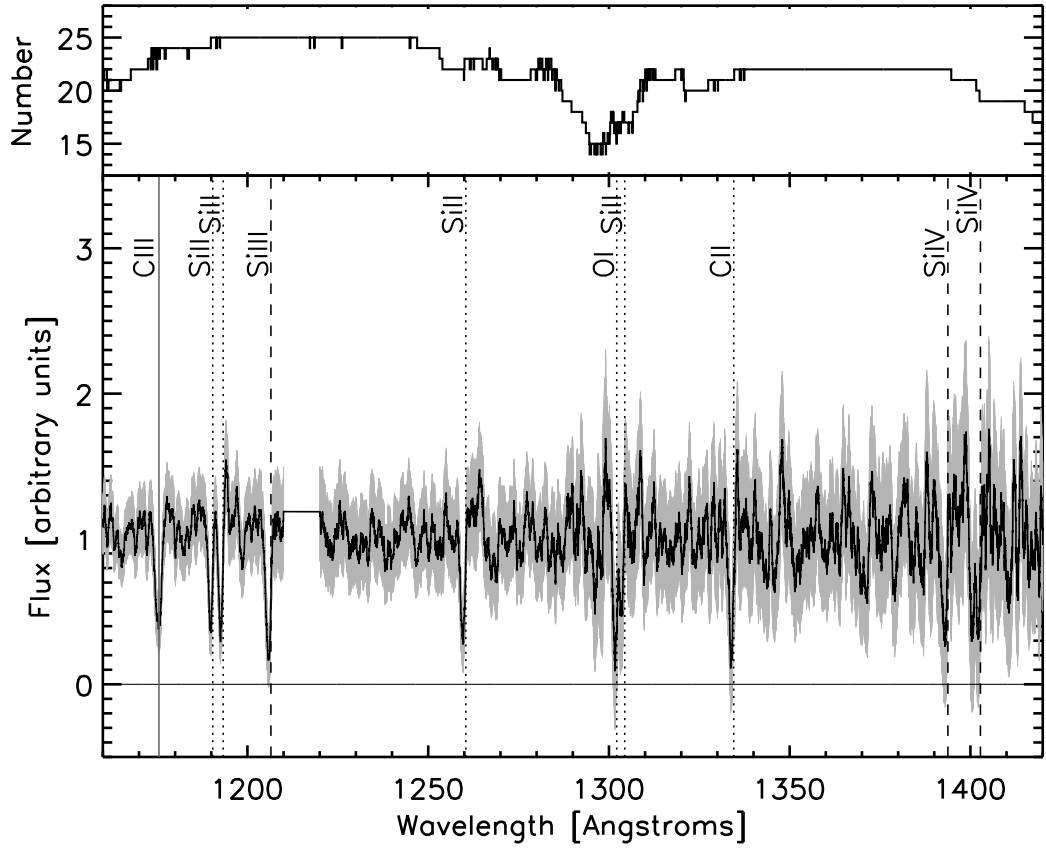}
  \caption{Composite rest-frame UV spectrum of 25 $z\sim 0.3$
    \lya-emitting galaxies. Multiple absorption features are
    identified with vertical lines. Solid lines indicate stellar
    photospheric absorptions, dotted and dashed lines indicates
    resonant absorption originating in the galaxy's ISM, with dotted
    and dashed showing low-- and high-- ionization metal lines,
    respectively. The top panel shows the number of galaxies used to
    compute the average spectrum at each wavelength.}
   \label{fig:stack}
\end{figure*}

To create the stacked spectrum, we first blueshift the observed
spectra of individual galaxies into the rest-frame using the measured
\ha\ velocities. Then, at each wavelength we compute a flux-weighted
average and a standard deviation. The mean stacked UV spectrum (shown
after a box-car smoothing of 0.85\AA) is shown in
Figure~\ref{fig:stack}, where the shaded gray area corresponds to
$\pm$ one weighted standard deviation. In Figure~\ref{fig:stack}, the
top panel shows the number of galaxies that entered the stack at each
wavelength. In the stacked spectrum we are able to clearly identify
and reliably measure the features presented in Tables~\ref{tab:abs}
and~\ref{tab:emi} and marked in Figure~\ref{fig:stack}. The list
includes five absorption lines and four fluorescent emission lines. The
list also includes the C~{\sc iii}$\,\lambda 1175$ absorption line,
which is mainly produced in the stellar photospheres. 

\begin{figure*}[t!]
   \centering
  \includegraphics[scale=0.4]{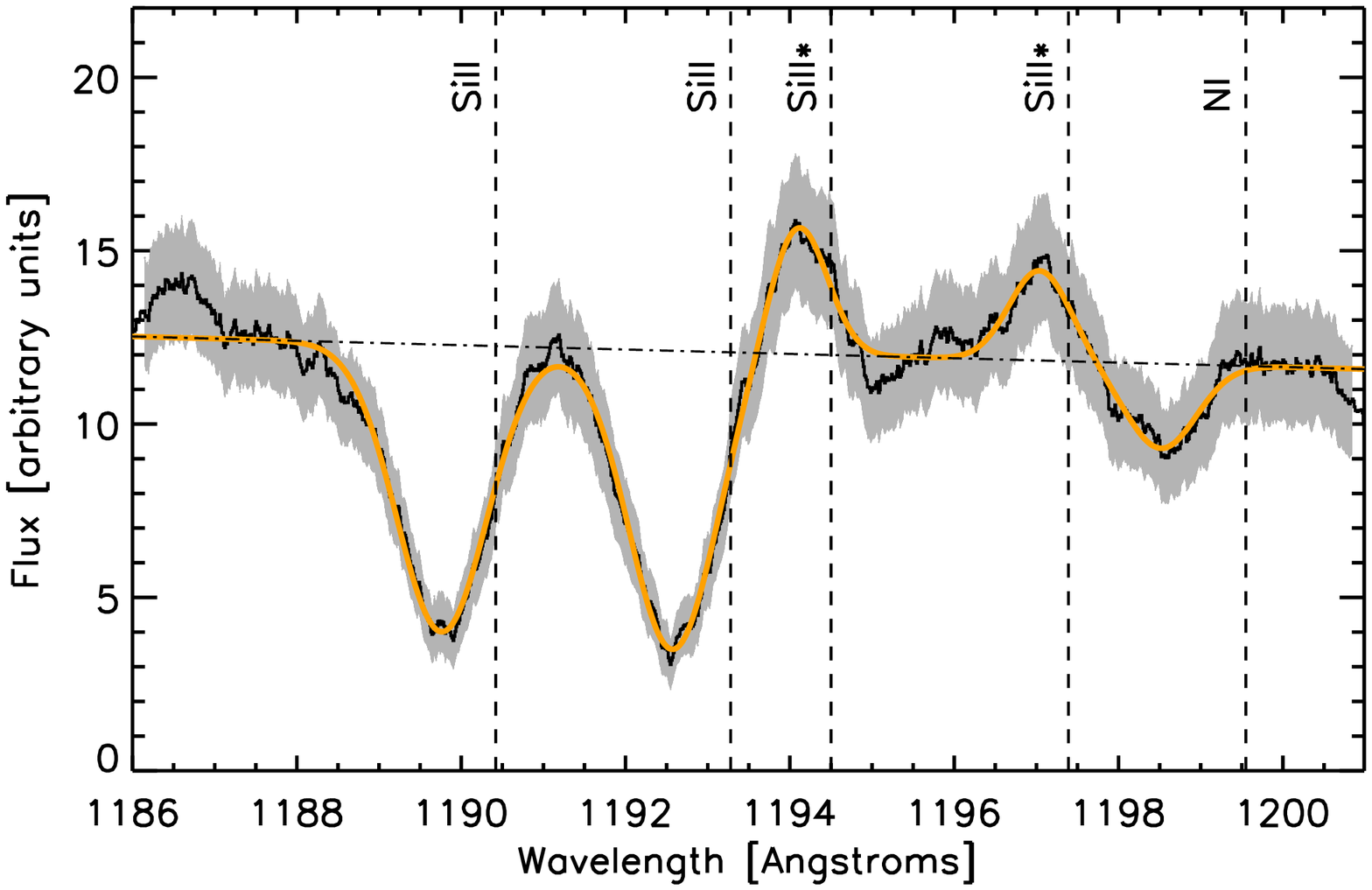}
  \includegraphics[scale=0.4]{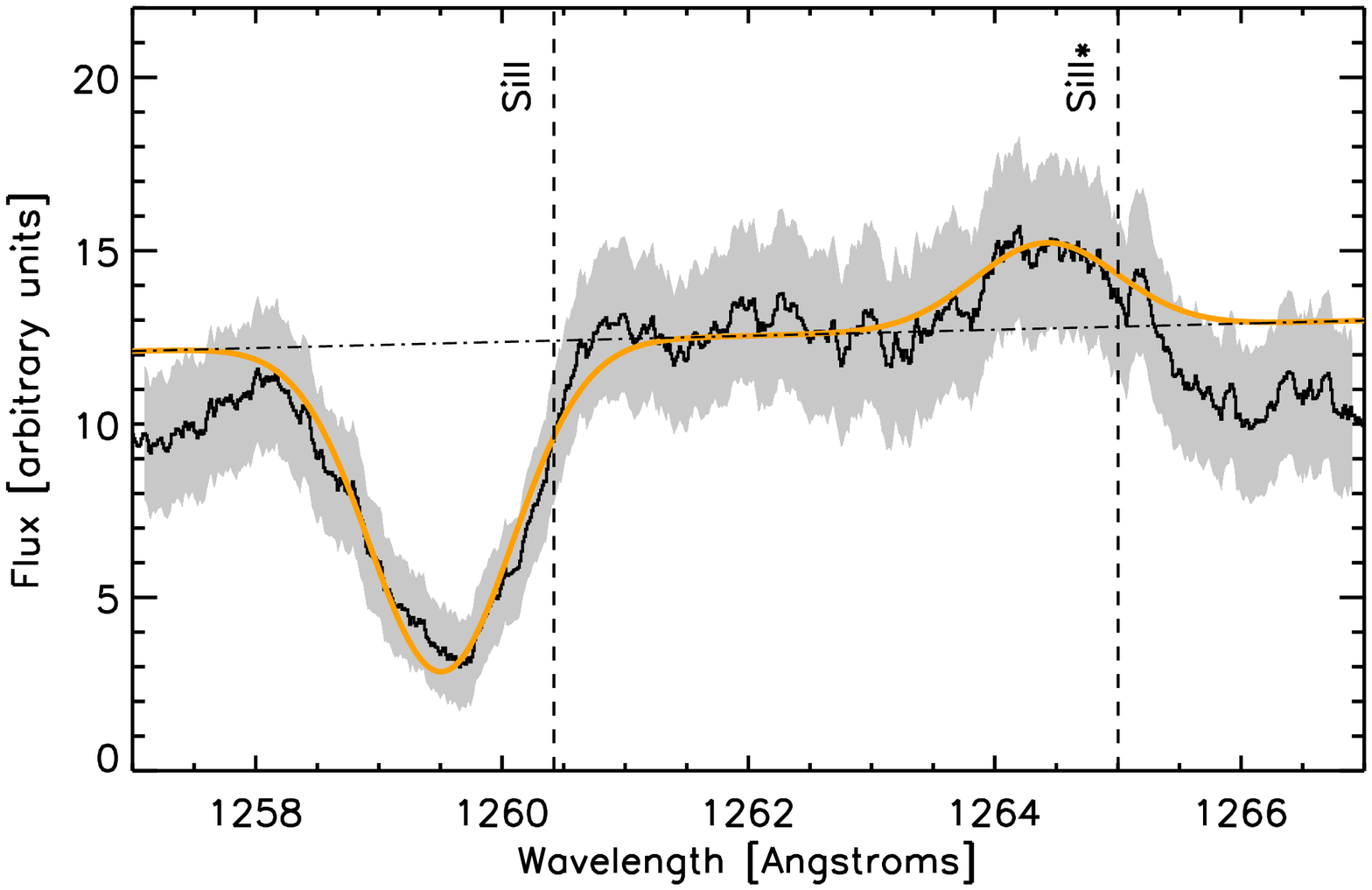}
  \caption{Zoom in on the spectral regions around the absorption and
    emission line features considered in this work. In each panel, the
    dashed vertical lines show the vacuum wavelength of each
    transition (as indicated by each label). The orange line shows the
    sum of the best-fit Gaussian profiles, while the dot-dashed line
    shows the continuum level.  }
   \label{fig:gaussian}
\end{figure*}

If we assume that lines are pure absorption and pure emission we could
measure the bulk velocity of the gas from the peak velocity of the
lines.  We derive the peak positions by fitting Gaussian line profiles
to the observed absorption/emission lines.  When two lines of a given
multiplet/ion are blended, we fit them simultaneously constraining the
width of the Gaussian function to be the same for both lines. In
Figure~\ref{fig:gaussian} we zoom--in on the spectral regions around
different transitions in the Si$^+$ and Si$^{++}$ ions and plot the
resulting best-fit Gaussian models. The errors on the peak wavelength
were computed with a Monte Carlo simulation. We created 1000
realizations of the stacked spectrum by changing the flux at each
wavelength within $\pm 1\,\sigma$. The new profiles were fitted with a
Gaussian, and the error on the peak wavelength was computed as the
standard deviation of the 1000 best--fit peak wavelengths.

In Tables~\ref{tab:abs} and \ref{tab:emi} we report the vacuum
wavelength of the considered transitions, the observed peak wavelength
of the profiles, as well as the velocity shift between the galaxy's
rest frame velocity (computed from the \ha) and the peak velocity of
the profiles. The stellar \ciii\ velocity is consistent with the
systemic velocity computed from the \ha\ emission line profiles.  Note
that the peak/trough velocities obtained from the Gaussian fits offer
an easy mathematical representation of the data but do not add
immediate physical meaning. The velocity profiles shown in
Figure~\ref{fig:gaussian} are typical of star forming galaxies at both
high and low redshifts
\citep[e.g.][]{shapley2003,steidel2011,jones2012,heckman2011,wofford2013},
where they are usually interpreted as originating in a gas outflow
probably driven by the current episode of star-formation.

The results of the Gaussian fits presented in
Figure~\ref{fig:gaussian} would indicate that the gas is moving toward
the observers with velocities --as measured at the maximum
absorption/emission-- ranging between $-160$ and $-220$\kms.  The
average velocity computed in this way from all absorption lines is
$-185 \pm 25$ \kms. The profiles in Figure~\ref{fig:gaussian} also
show absorption at velocities as high as $-500$\kms, indicating the
presence of multiple velocity components and/or a velocity gradient in
the outflowing gas. All detected fluorescent emission lines are
blueshifted with respect to the systemic \ha\ velocity with peak
velocities ranging between $-88$ and $-137$\kms. With an average
outflow velocity of $-100\pm 22$\kms, the fluorescent emission
components appear to have a systematically-lower velocity shift than
the resonant absorption lines.

Resonant photons can be either
re--emitted at resonance, or in the fluorescent channel, when
available. Thus, the two lines originate in the same gas and will
share the same kinematical properties. Naively, the measured
systematic difference in the bulk velocities of the absorption and
fluorescent emission lines could then be interpreted as an indication
that these lines formed in two kinematically--distinct components.  In
the following section, we use SALT to consistently model the
scattering from the outflowing gas, and show how the systematic
velocity difference can be the result of a non-symmetric outflow.

\subsection{Modeling the line profiles}
\label{sec:data_modeling}
Here we use SALT to model the observed absorption/emission profiles
observed in the stacked spectrum. The free parameters for the models
are $\tau_0$, $v_0$, and $v_{\infty}$ for the spherical outflow, and
$\tau_0$, $v_0$, $v_{\infty}$ and $R_{\rm aper}$, for the spherical
outflow plus aperture. These parameters fully describe the density and
velocity field of the galactic outflow, and do not depend on the
particular transition within a given ionic species.  We therefore
constrain the model's free parameters by simultaneously fitting the
four radiative transitions of Si$^+$, observed around
$\lambda=1190$\AA. We chose this spectral region because it shows the
highest $S/N$ of the stacked spectrum, and because of the presence of
both two resonant absorption and the corresponding fluorescent
emission lines. To model the observed line profiles we use
equation~\ref{eqn:fullMS}, which accounts multiple scattering within
each shell as explained in section~\ref{sec:model}.

\begin{figure}[t!]
   \centering
  \includegraphics[scale=0.5]{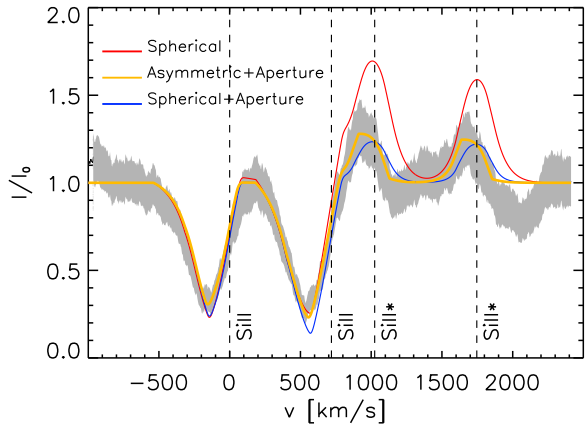}
 \caption{The absorption and re--emission profiles of the Si~{\sc ii}
    doublet are well reproduced with an asymmetric model and
    accounting for the effect of the finite COS aperture size. Best
    fit model profiles to the Si~{\sc ii}~1190--1193\AA\ doublet are
    shown for different outflow geometries, as indicated in the
    label. All models shown in this Figure include multiple scattering
    within each shell.}
   \label{fig:model_fits}
\end{figure}

We derive the best-fit parameters for the symmetric outflow model
with and without a view-limiting aperture, by performing a $\chi^2$
minimization on the entire doublet profile, including both absorption
and emission lines. The models were computed on the same velocity
vector as the data, and then box-car smoothed with the same kernel,
before proceeding to the computation and minimization of the $\chi^2$.
The best--fit line profiles for the SiII doublet are shown in
Figure~\ref{fig:model_fits}, and the parameter values are given in
Table~\ref{tab:model_fits}. The blue and red curves show the best fit
for the spherical outflow with and without the spectroscopic aperture,
respectively.

The simplest spherical models (with or without aperture) well
reproduce the depth, shape and central velocities of the blueshifted
absorption components of the resonant doublet. This indicates that our
simple assumptions for the density and velocity fields of the
scattering material are adequate representation of the $\rm Si^+$
distribution. On the other hand, the spherical model with no  limiting aperture
fails to reproduce the observed profiles in two key aspects: 1) it
substantially overproduces the amount of scattered emission, and
2), due to the symmetry in the considered configuration, it
predicts that the emission component of the P-Cygni profile should be
centered at the systemic velocity, while the observations shows that
the peaks of the fluorescent emission are clearly blueshifted.

As we discussed in Section~\ref{sec:aperture}, the effect of a
spectroscopic aperture is to selectively decrease the number of
scattered photons that are able to reach the observer. As
Figure~\ref{fig:model_fits} shows (blue curve), adding an aperture
alleviates the first of the two discrepancies. However, because of the
intrinsic symmetry of the outflow model, the scattered re--emission is
still centered at zero velocity, and therefore the model still fails
in fully reproducing the observed features.  A blueshift in the
scattered emission component originating in outflowing material can be
obtained if the outflow is not spherically symmetric with respect to
the central source. A simple way to implement this, is by
differentially weighting the contribution to the final profile from
different portions of the envelope. Thus, we introduce a scaling
factor -- $f_{\rm obsc}$ -- to the red component of the scattered
profile (Eqn~\ref{eqn:red}); i.e., the radiation scattered in the half
sphere moving away from the observer).  Physically, the parameter
$f_{\rm obsc}$ can be used to mimic a face-on galaxy, where the disk
is absorbing part of the radiation emitted by the outflowing material
(see Section~\ref{sec:discussion}).  The profile can now be written
as:

\begin{equation}
I(x)^{asy}=1-I_{\rm abs} + I_{\rm emi, blue} + f_{\rm obsc} \times
I_{\rm emi, red}.
\end{equation}

\noindent
$f_{\rm obsc}$, represents the fraction of $I_{\rm emi, red}$ that is
allowed to reach the observer. We refer to the above profile as the
``asymmetric model'', to indicate that the receding half of the
expanding envelope is seen less easily than the approaching front
part.  The best fit asymmetric outflow model is shown in
Figure~\ref{fig:model_fits} with the yellow line. This model is
clearly able to \textit{simultaneously reproduce the relative
  intensity of the fluorescent emission and resonant absorption lines,
  as well as their systematically different peak positions}.

\begin{figure}[h]
   \centering
  \includegraphics[scale=0.5]{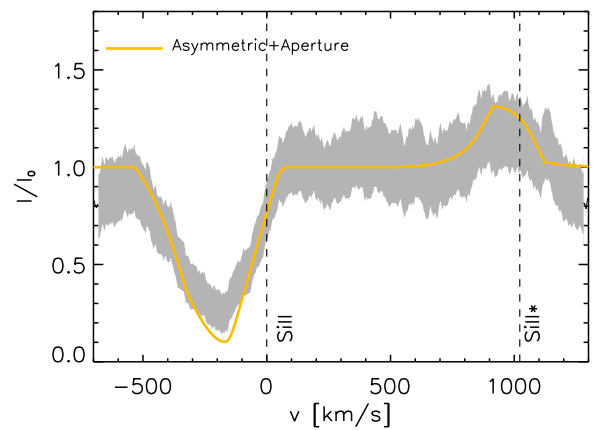}
 \caption{ Si~{\sc ii}$\lambda 1260$ absorption line profile and
    corresponding fluorescent emission, together with the models
    computed using the best-fit outflow parameters derived from the
    \siii~1190--1193\AA\ doublet.}
   \label{fig:siii1260}
\end{figure}

We can test our results using a different resonant transition in
Si$+$. The resonant line at $\lambda=1260.42$ is ideal for this
purpose. As Figure~\ref{fig:gaussian} shows, we detect both the
resonant absorption and the corresponding fluorescent emission. This
absorption originates in the same material where the Si~{\sc
  ii}$\lambda 1190$ doublet is produced and is thus perfect to test
the parameters of the outflow model (i.e., the density, velocity
field, and geometry). In Figure~\ref{fig:siii1260} we show the
observed Si~{\sc ii}$\lambda 1260$ profile, together with the model
profile computed using the best-fit outflow parameters derived from
our analysis of the 1190--1193\AA\ doublet (i.e., changing only the
transition dependent parameters in the profile
equations). Figure~\ref{fig:siii1260} shows that the model optimized
to fit the \siii\ doublet fully reproduces the observed Si~{\sc
  ii}$\lambda 1260$ profile as well. In particular, it reproduces both
the blueshifted absorption component as well as the intensity and peak
wavelength of the fluorescent emission. We stress again that the
parameters of the outflow are kept fixed to the best-fit values given
in Table~\ref{tab:model_fits}.

\section{Discussion}
\label{sec:discussion}
Near and far--UV spectra include numerous resonant metal absorption
lines that, combined with the appropriate theoretical tools, provide
powerful diagnostics for galactic outflows. High--quality rest--frame
UV spectra are currently available for nearby individual galaxies
(e.g., with data from the Hubble Space Telescope), and stacked spectra
of high--redshift galaxies (with data from 8m class telescopes). Soon,
with the planned 30m telescopes, we will be able to study at high
resolution absorption line spectra in individual objects up to the
highest redshifts.  

Resonant blue--shifted UV absorption features are commonly modeled as
originating in a thin shell of gas moving at the outflow velocity
\citep[as measured from the centroids of resonant absorption lines,
e.g.,][]{verhamme2006,schaerer2008,verhamme2008,schaerer2011}. Substantial
evidence, however, suggests that this description is too simplistic,
as noted already by, e.g., \citet{pettini2002}. First of all, when the
absorption lines are observed at high enough spectral resolution, they
show asymmetric profiles covering a broad range of velocities
\citep[up to as much as -1000 \kms,
][]{tremonti2007,diamondstanic2012,sell2014}. This indicates that the
gas is not confined to a thin shell, but rather is distributed in an
extended envelope, with velocity and density changing with the
distance from the galaxy.  More realistic models of extended outflows,
with velocity and density gradients have been proposed
\citep[e.g.,][]{prochaska2011,rubin2011,martin2013}, and highlight the
importance of properly accounting for the geometry of the outflowing
material.

Second, well defined P--Cigny profiles from resonant transitions of
MgII, as well as the detection of fluorescent emission associated with
resonant transitions of \siii, and FeII, are commonly observed
\citep[][]{shapley2003,france2010,rubin2011}, indicating that the
re--emission component from the outflowing gas cannot be neglected,
particularly in compact galaxies or galaxies at high--z, where the
spectroscopic aperture may include a substantial fraction of the
extended scattering outflow. Neglecting possible contribution from
re--emitted photons may have important consequences for the
determination of the gas column density and/or covering fraction, as
recently noted by, e.g., \citet{prochaska2011}.

The SALT model discussed in this paper provides a simple analytical
description of the line profiles originating in the expanding
envelopes around galaxies, that properly account for multiple
scattering of resonant photons, scattered re-emission, and
observational aperture effects.

\subsection{On the use of the absorption profiles as indication of
  covering fraction}
In the approximation of pure absorption and with well resolved line
profiles, the gas apparent optical depth at a given velocity $\tau(v)$
is often used to derive the apparent column density profile
\citep[$-ln(I(v)/I_0(v))=\tau(v) \propto f\lambda N(v)$,
e.g.][]{pettini2002}. However, it is well known that the apparent
column density obtained from line profiles can be underestimated if
undetected saturated components contribute at some velocities. When
two or more transitions of a given species differing only in the
product of $f \lambda$ are available, then information about line
saturation can be inferred from the comparison of the apparent optical
depth profiles (that should be identical within the observational
errors). If saturation is present, the line with the highest
oscillator strength will result in a lower apparent column
density. The previous reasoning is correct only if the absorbing gas
fully covers the continuum source. If this is not the case, the line
profile will also depend on the gas covering fraction ($f_C$), {\it as
  well as} the optical depth (i.e., $I/I_0=1-f_C (1-e^{- \tau } )$).
Various works have used absorption-line profiles to determine the gas
$f_C$, under the assumption that absorption lines are saturated
\citep[i.e., $e^{-\tau} \rightarrow 0$ in $I/I_0$,
e.g.][]{jones2013,martin2009}.  When the scattering envelope is
included in the spectroscopic aperture, however, this approach cannot
be used: the resonant scattered re--emission affects substantially the
residual intensity at line center, in different ways depending on the
atomic level structure and the spontaneous transition probabilities..

\begin{figure}[h]
  \centering
\includegraphics[scale=0.5]{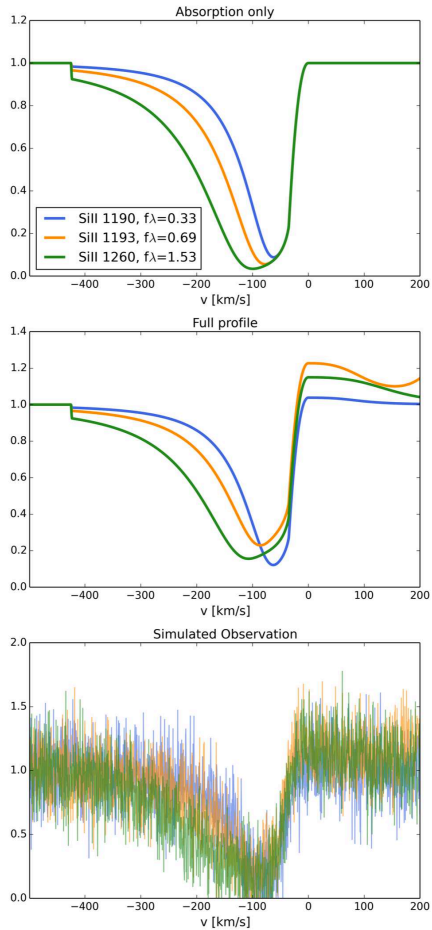}
\caption{Neglecting emission filling of the absorption profiles can
  cause erroneous conclusions regarding the gas covering fraction.
  Top: absorption-only components of three resonant transitions of
  Si$+$, as indicated in the legend. The model correspond to a
  spherical envelope with $v_0=35$km/s, $v_{\infty}=450$km/s, and
  $\tau_{1190\AA}=30$.  When only the absorption components are
  considered, the profiles scale according to the value of
  $f\lambda$. Middle: full line profiles including the scattered
  re--emission component.  Particularly at velocities close to zero,
  the profiles do not scale as $f\lambda$ anymore. Bottom: simulated
  spectrum with 10\% errors, and resolution of 30km/s.}
  \label{fig:discussion}
\end{figure}

We show this point in Figure~\ref{fig:discussion}, where we present
the resonant absorption profiles originating in an outflowing envelope
with $f_c=1$, for three transitions of Si$^+$. The top panel shows the
absorption component only of the profile: when only absorption is
considered, the three profiles scales as expected, according to the
value of $f\lambda$ for each transition. This is not true anymore when
the resonant scattered and fluorescent re-emission are included
(middle panel): close to the maximum absorption, in fact, the
transmission with the lowest value of $f\lambda$ is in fact the
deepest! This is simply due to the relative value of the branching
ratios for the resonant and fluorescent channels for these particular
transitions.  At the largest velocities, where the filling from
scattered radiation is less important and the optical depth is
smaller, the absorption profiles are again proportional to
$f\lambda$. 

In the bottom panel of Figure~\ref{fig:discussion} we simulate the
line profiles as they would be observed with COS, assuming a spectral
resolution of 30\kms, and a noise of 10\%.  Because of the resolution
and S/N ratio, the three lines are identical close to the core, and
barely distinguishable at the largest velocities. We note, however,
that we did not account for uncertainties in the normalization of the
spectra, which may affect the profiles, particularly at the largest
velocities. The simulated lines also show that from the resonant
profiles alone it is hard to identify the presence of scattered
re--emission: because at least 50\% of the photons are re--emitted in
the fluorescent line, for all the transitions considered. However,
\textit{the presence of the fluorescent emission is a clear sign--post
  that the absorption profiles will be affected by emission
  filling}. Because the strength of the filling depends on the
branching ratio for the fluorescent transitions, it is not correct to
average together profiles of different lines of the same
ion. Moreover, the effect of the filling of the absorption lines will
also depend on the {\it geometry of the outflow} (see
Section~\ref{sec:data_modeling}) as well as the {\it the relative size
  of the scattering outflow region and the spectroscopic aperture.}.

As an example of what can be achieved by modeling absorption profiles
with SALT, we have modeled the UV stacked spectrum of 25 \lya\ emitters
in the nearby Universe. The spectrum shows a number
of resonant absorption and fluorescent emission lines generating in
the $\rm Si^+$ in the galaxy's gaseous medium. The profiles of
the resonant absorption lines are systematically blueshifted with
respect to the systemic velocity, with an average gas velocity at
maximum absorption of $-185 \pm 25$ \kms. The peak velocity of the
fluorescent emission lines, however, is systematically lower than the
outflow velocity obtained from the absorption profiles ($-100\pm 22$
\kms). We showed with SALT how simple symmetric models fail to
reproduce this velocity difference, and asymmetry in the gas
distribution needs to be present. Our best fit model with
$f_{obsc}=0.1$ describes a geometry in which the half a sphere
receding from the observer contributes only $10$\% to the fluorescent
emission, which is then dominated by the emission from the half a
sphere approaching the observer. Physically, this simple model can be
used to describe the outflow from a disk galaxy seen approximately
face--on, where the radiation scattered from the half--sphere receding
from the observer has to go through the opaque disk, that will have an
optical depth $\tau = -ln(1-f_{\rm obsc})=2.3$.

The sample of galaxies that entered the stacked analysis is not
randomly selected among starforming objects, but rather it comprises
only galaxies showing \lya\ in emission. The conditions that allow
more \lya\ radiation to escape from a galaxy are still highly debated,
with some authors suggesting that age is a dominant factor
\citep[e.g.,][]{cowie2011}, while others advocating for dust
extinction \citep[e.g.][]{hayes2013,atek2014}. Our result seems to
indicate that, on average, \lya-emitters are able to escape more
easily in the direction perpendicular to the galaxy disk, than along
the disk, suggesting that the viewing angle is also an important
factor in determining the \lya\ escape/visibility. This result is not
unexpected, since the direction perpendicular to the galactic disk is
also the one that offers the minimum column density of diffuse
material, and, thus, of neutral gas \citep[e.g., see ][for results of
radiative transfer simulations of \lya\ photons in a realistic spiral
galaxy]{verhamme2012}. In support of our findings, a recent structural
analysis of hundreds of \lya\ emitters at $z\sim 2.2$ showed that
\lya\ emitters tend to have smaller ellipticity than galaxies at
similar redshift with no \lya\ emission, reinforcing the idea that
\lya\ photons escape more easily in the direction of minimum optical
depth, i.e., minimum hydrogen column density \citep{shibuya2014}.

\section{Conclusions}
In this work we have presented and discussed a simple Semi--Analytical
Line Transfer model, SALT, to describe the expected absorption and
re--emission line profile generated in a spherical outflow surrounding
a galaxy with a finite size ($R_{SF}$), under the Sobolev
approximation. We derive the analytical profiles computed for the
velocity field with the velocity increasing with distance from the
galaxy ($v\propto r^{\gamma}$), and the densitity radial profile
obtained assuming a constant mass outflow rate
($n_l(r)\propto (vr^2)^{-1}$). We include the effect of multiple
scatterings properly accounting for the atomic structure of the
scattering ions. We also discuss how the line profile changes due to
the effect of a circular spectroscopic aperture that does not cover
the full extent of the outflow
(Section~\ref{sec:comparison_observations}), and the case in which
part of the outflow is obscured.

Our analysis reproduces the main features observed in the profiles
generated with more complex radiative transfer codes applied to
gaseous outflows with the same geometry \citep{prochaska2011}.
Namely, we show how -for a spherically symmetric outflow- the
scattered re--emission is centered at zero velocity and thus alters
the shape of the pure absorption component generated in the material
in front of the emitting galaxy. Outflow velocities computed from the
wavelength of the absorption line trough do not include this
effect. Such an analysis results in an overestimate of the outflow
velocity.  The intensity of the emission component of the line profile
depends not only on the spectroscopic aperture used for the
observations, but also on the atomic structure of the particular ion
used in the analysis and the spontaneous transition probabilities. In
the case of the Si$^+$1190--1193\AA\ doublet, a photon absorbed in the
1190\AA\ transition has a higher chance of being re--emitted in the
fluorescent transition to the 2$P^0 3/2$ level, thus somewhat limiting
the ``filling'' effect on the absorption component.

We have considered the resonance absorption and fluorescent emission
profiles observed in the average UV spectrum of 25 $z\sim 0.3$ \lya\
emitters. With a simplistic Gaussian profile fit one would find that
the average velocities computed from the trough ($-185 \pm 25$ km
s$^{−1}$) and the peak wavelengths ($-100 \pm 22$ km s$^{−1}$) of the
profiles systematically differ, with the absorption-derived velocity
being more negative than the emission-derived velocity.  Regardless of
the size of the spectroscopic aperture, a symmetric outflow of an
arbitrary shape produces a profile with the emission component
centered at the systemic velocity of the galaxy, and so cannot explain
the systematic difference between the emission and absorption
velocities.  On the other hand, we used SALT to show that this shifts
comes naturally if most of the radiation scattered by the receding
half of the outflow is obscured from view.  This model can be
interpreted as a simple representation of a disk galaxy observed
face--on, where the thick disk acts as a semi--transparent screen for
the backscattered radiation.  This result thus indicates that, on
average, galaxies tend to show \lya\ in emission more frequently when
observed face--on. This idea is supported both by recent observations
of high--redshift \lya--emitting galaxies \citep{shibuya2014}, as well
as high--resolution radiative transfer simulation of \lya\ photons in
disk galaxies \citep{verhamme2012}.

To conclude, by simultaneously reproducing both the resonant
absorption and the associated resonant and fluorescent emission, the
line profiles computed with our SALT model are more far-reaching than
a simple absorption--line based analysis. This is especially true when
the data show evidence (e.g., the presence of a P-Cygni profile and/or
fluorescent emission lines) of scattered re--emission from the galaxy
wind. The formalism developed here can be easily extended to other
geometries, to account for clumpiness of the outflowing gas (Card et
al., in prep), and to different ions/transitions.

\acknowledgments We wish to thank our referee for valuable comments
that helped us to improve the presentation of our work. CS
acknowledges Alaina Henry, Crystal Martin, Dawn Erb, Marc Dijkstra for
stimulating discussions.  CS acknowledges partial support by
HST-GO-12269.01 grant. NP acknowledges partial support by STScI--DDRF
grant D0001.82435. CS acknowledges M. Bagley for a careful reading of
the manuscript (and pointing out the randomness in the use of commas).

\begin{deluxetable*}{cccccccc} 
  \tablecaption{Atomic data for \siii\ and \siiii\ ions. Data taken
    from the NIST Atomic Spectra
    Database\footnote{http://www.nist.gov/pml/data/asd.cfm}.\label{tab:siII}}
  \tablehead{
    \colhead{Ion} &\colhead{Vac. Wavelength} & \colhead{$A_{ul}$} & \colhead{$f_{lu}$} &\colhead{$E_{l}-E_{u}$}&\colhead{$g_l-g_u$} & \colhead{Lower level} & \colhead{Upper level}\\
    \colhead{} &\colhead{\AA} & \colhead{s$^{-1}$}
    &\colhead{}&\colhead{eV}&\colhead{}&\colhead{Conf.,Term,
      J}&\colhead{Conf.,Term, J}} \startdata
  \siii\ &1190.42&6.53$\times10^8$&  	 2.77$\times 10^{-1}$ &$0.0 - 10.41520$&$2-4$& $3s^23p \,  2P^0  \,  1/2$  &	 $3s3p^2  \, 	 2P  \, 	 3/2$\\
  &1193.28&2.69$\times10^9$&  	 5.75$\times 10^{-1}$ &$0.0 - 10.39012$&$2-2$& $3s^23p \,    2P^0 \,  1/2$  &	 $3s3p^2  \, 	 2P  \, 	 1/2$\\
  &1194.50&3.45$\times10^9$&  	 7.37$\times 10^{-1}$ &$0.035613 - 10.41520$&$4-4$& $3s^23p \,   2P^0 \,  3/2$  &	 $3s3p^2  \, 	 2P  \, 	 3/2$\\
  &1197.39&1.40$\times10^9$&  	 1.50$\times 10^{-1}$ &$0.035613 - 10.39012$&$4-2$& $3s^23p \,    2P^0 \,   3/2$  &	 $3s3p^2  \, 	 2P  \, 	 1/2$\\
  &1260.42 & 	2.57$\times 10^9$ & 	 1.22 	 &        $  0.0    	- 	 9.836720 $ &  $ 2-4$ &		 $3s^23p \,	 2P^0 \,	 1/2 $&	 $3s^23d \,	 2D 	\, 3/2 $\\
  &1264.73 &	3.04$\times 10^9$ &	 1.09 	 &      $0.035613   - 	 9.838768$ &   $4-6$  &	 $3s^23p \,	 2P^0\, 	 3/2$& $	 3s^23d \,	 2D\, 	 5/2$\\
\enddata 
\end{deluxetable*}

\begin{deluxetable}{llccc} 
\tablecaption{Absorption lines measured in the stacked spectrum.\label{tab:abs}}
\tablehead{ 
\colhead{Species} & Formation & \colhead{$\lambda_\mathrm{vac}$} &
\colhead{$\lambda_{\mathrm{obs}}$} &  \colhead{$\Delta v$ \tablenotemark{a} } \\
\colhead{} & &  \colhead{\AA}   &  \colhead{\AA} }
\startdata 
Si~{\sc ii}   & ISM   & 1190.42 & 1189.83 $\pm 0.09$& --160 $\pm$ 25\\ 
Si~{\sc ii}   & ISM   & 1193.29 & 1192.60 $\pm 0.09$ & --174  $\pm$ 24\\ 
Si~{\sc ii}   & ISM   & 1260.42 & 1259.56$\pm 0.08$ & --218 $\pm 21$\\
C~{\sc ii}  & ISM   & 1334.53 & 1333.81 $\pm 0.10$ & --164$\pm$  24\\
Si~{\sc iii}  & ISM   & 1206.50 & 1205.75 $\pm 0.08$ & --188$\pm 20$  \\
C~{\sc iii}   & Photo & 1175.53 & 1175.42$\pm0.1$ & --28 $\pm$25

\enddata 
\tablenotetext{a}{Velocity shift of the absorption trough with
  respect to the \ha\ emission line systemic velocity. }
\end{deluxetable}

\begin{deluxetable}{llccc} 
\tablecaption{Fluorescent emission lines measured in the stacked spectrum.\label{tab:emi}}
\tablehead{ 
\colhead{Species} & Formation & \colhead{$\lambda_\mathrm{vac}$} &  \colhead{$\lambda_{\mathrm{obs}}$} &  \colhead{$\Delta v$ \tablenotemark{a} } \\
\colhead{} & &  \colhead{\AA}   &  \colhead{\AA} }
\startdata 
Si~{\sc ii}$^*$   & ISM   & 1194.50 & 1194.10 & --100.2 \\ 
Si~{\sc ii}$^*$   & ISM   & 1197.39 & 1197.04 & --88.2\\ 
Si~{\sc ii}$^*$   & ISM   & 1265.00 & 1264.42 & --137.5\\
 C~{\sc ii}$^*$   & ISM   & 1335.71 & 1334.53 & --55.7 
\enddata 
\tablenotetext{a}{Velocity shift of the absorption trough with
  respect to the \ha\ emission line systemic velocity. }
\end{deluxetable}

\begin{deluxetable*}{lccccc} 
\tablecaption{Best fit values for the \siii\, $lambda 1190-1193$ doublet.\label{tab:model_fits}}
\tablehead{ 
\colhead{Model} &\colhead{$v_0$} & \colhead{$v_{\infty}$} & \colhead{$\tau_0$} &
\colhead{$\frac{R_{\rm aper}}{R_{\rm SF}}$}&\colhead{$f_{\rm obsc}$}\\
\colhead{}&\colhead{[\kms]} & \colhead{[\kms]} & \colhead{} & \colhead{}&\colhead{} 
}

\startdata 
Spherical model full view & 38&425&160&\dots&\dots\\
Spherical model -- limited view& 40&426&120&2&\dots\\
Asymmetric model -- limited view & 55&425&45&3&0.1
\enddata 
\end{deluxetable*}

\bibliographystyle{apj} 
\bibliography{myreferences}{}

\end{document}